\documentclass[aps,prl,twocolumn,amsfonts,longbibliography,citeautoscript,superscriptaddress]{revtex4-1}

\usepackage{color,bm,bbold}
\usepackage{amsmath,amssymb}
\usepackage{epsfig,array}
\usepackage{verbatim}
\usepackage[T1]{fontenc}

\newcommand{\be}{\begin{equation}}
\newcommand{\ee}{\end{equation}}
\newcommand{\bw}{\begin{widetext}}
\newcommand{\ew}{\end{widetext}}
\newcommand{\la}{\langle}
\newcommand{\ra}{\rangle}
\newcommand{\dg}{^\dagger}
\newcommand{\p}{\partial}
\newcommand{\rd}{{\rm d}}
\newcommand{\s}{\sigma}
\def\nn{\nonumber\\}
\def\fr#1{(\ref{#1})}

\usepackage[colorlinks=true,bookmarks=false,citecolor=blue,linkcolor=red,hyperfootnotes=true,urlcolor=blue]{hyperref}

\begin{document}
\title{Nonthermal states arising from confinement in one and two dimensions}

\author{Andrew J. A. James}
\email{andrew.james@ucl.ac.uk}
\affiliation{London Centre for Nanotechnology, University College London, Gordon Street, London WC1H 0AH, United Kingdom}

\author{Robert M. Konik}
\email{rmk@bnl.gov}
\affiliation{Condensed Matter Physics \& Materials Science Division, Brookhaven National Laboratory, Upton, NY 11973-5000, USA}

\author{Neil J. Robinson}
\email{n.j.robinson@uva.nl}
\affiliation{Institute for Theoretical Physics, University of Amsterdam, Science Park 904, 1098 XH Amsterdam, The Netherlands}

\begin{abstract}
We show that confinement in the quantum Ising model leads to nonthermal eigenstates, in both continuum and lattice theories, in both one (1D) and two dimensions (2D). In the ordered phase, the presence of a confining longitudinal field leads to a profound restructuring of the excitation spectrum, with the low-energy two-particle continuum being replaced by discrete `meson' modes (linearly confined pairs of domain walls). These modes exist far into the spectrum and are atypical, in the sense that expectation values in the state with energy $E$ do not agree with the microcanonical (thermal) ensemble prediction. Single meson states persist above the two meson threshold, due to a surprising lack of hybridization with the ($n\geq4$)-domain wall continuum, a result that survives into the thermodynamic limit and that can be understood from analytical calculations.  The presence of such states is revealed in anomalous post-quench dynamics, such as the lack of a light cone, the suppression of the growth of entanglement entropy, and the absence of thermalization for some initial states.  The nonthermal states are confined to the ordered phase -- the disordered (paramagnetic) phase exhibits typical thermalization patterns in both 1D and 2D in the absence of integrability.
\end{abstract}

\maketitle

\noindent {\bf Introduction:}
In a generic quantum many-body system eigenstates are thermal, in the sense that expectation values (EVs) within an eigenstate agree with the microcanonical ensemble (MCE, thermal) prediction~\cite{deutsch1991quantum,srednicki1994chaos,srednicki1999approach,rigol2008thermalization,rigol2009breakdown,rigol2012alternatives,santos2010localization,santos2010onset,ikeda2013finitesize,kim2014testing,beugeling2014finitesize,sorg2014relaxation,dalessio2015from,khodja2015relevance,cosme2015relaxation,mondaini2016eigenstate,modaini2017eigenstate,lan2017eigenstate}. At the heart of understanding this behavior is the eigenstate thermalization hypothesis (ETH)~\cite{deutsch1991quantum,srednicki1994chaos,dalessio2015from}, which proposes a simple set of criteria for this to occur. Briefly summarized, the ETH tells us that an operator $\hat O$ will have a thermal EV in an eigenstate $|\alpha\ra$ with energy $E_\alpha$ provided its matrix elements satisfy~\cite{srednicki1999approach,dalessio2015from}
\be
\hat O_{\alpha,\beta} = \bar O(E) \delta_{\alpha,\beta} + e^{-S(E)/2} \breve O(E,\omega) R_{\alpha,\beta}. \label{eth}
\ee
Here $E = (E_\alpha + E_\beta)/2$, $\omega = (E_\alpha-E_\beta)$, and the ETH supposes $\bar O(E)$ is such that $\hat O_{\alpha+1,\alpha+1} - \hat O_{\alpha,\alpha} \propto e^{-R}$ (with $R$ the system size). The off-diagonal elements are suppressed by the thermodynamic entropy $S(E)$ and characterized by a well-behaved smooth function $\breve O(E,\omega)$ and a random variable $R_{\alpha,\beta}$ with zero mean and unit variance.

It is generally assumed that a non-integrable quantum many-body system obeys the ETH, with matrix elements of local observables satisfying Eq.~\fr{eth} (see, e.g, the short argument in~\cite{modaini2017eigenstate}). In large volume, at finite energies, the extensivity of $S(E)$ suppresses the second term in~\fr{eth}, and EVs are governed solely by the smooth function $\bar O(E)$. EVs are then thermal by construction, as the MCE prediction coincides with $\bar O(E)$. In a finite volume, there is some variance about $\bar O(E)$, which shrinks to zero with increasing system size. 

It is known, however, that nonthermal states that violate the ETH can also exist in finite size systems~\cite{biroli2010effect,santos2010localization,santos2010onset,richter2018realtime,robinson2018signatures}, usually being observed at the very edges of the spectrum. The presence of such nonthermal states in the spectrum can have important consequences for nonequilibrium dynamics~\cite{polkovnikov2011nonequilibrium,gogolin2015equilibration,dalessio2015from,richter2018realtime,robinson2018signatures}, in particular leading to an absence of thermalization following a quantum quench~\cite{biroli2010effect,robinson2018signatures}. Thermalization is used here in the sense that EVs  in the diagonal ensemble (DE) agree with the microcanonical ensemble (MCE) result~\cite{rigol2008thermalization,rigol2009breakdown}. Such predictions can now be routinely tested in cold atomic gases, following ground breaking progress in isolating and controlling such systems~\cite{weiss2006quantum,lewenstein2007ultracold}. 

In this Letter, we show that nonthermal states can exist away from the very edges of the spectrum in theories with confinement, both in 1D and 2D. We will show this to be true both on the lattice and in the scaling (continuum) limit. In the continuum limit, which is not usually the subject of ETH studies, we are able to harness powerful numerical techniques~\cite{james2018nonperturbative} to look at large system sizes, and present systematic analytical calculations that support our results. On the lattice, we use matrix product state (MPS) methods~\cite{schollwock2011densitymatrix} to show that the observed physics is not a remnant of the scaling limit and so may be possible to probe in experiments on low-dimensional quantum magnets (see, e.g.,~\cite{coldea2010quantum,morris2014hierarchy,wang2015spinon,wang2016from}). 

\noindent {\bf 1D lattice and continuum theories:}
Let us focus on a particular example of a theory with confinement, the quantum Ising chain with an additional longitudinal field 
\be
H_\text{latt} =  \sum_{j=1}^N J \s^z_j \s^z_{j+1} + h_x \s^x_j + h_z \s^z_j.  \label{lattice1D}
\ee
Here $\s^\alpha_j$ ($\alpha=x,y,z$) are the Pauli matrices acting on the $j$th site of the chain, $J$ is the Ising exchange parameter, and $h_x$ ($h_z$) is the transverse (longitudinal) field strength. Taking the scaling limit in the vicinity of the critical point ($h_x = 1$, $h_z = 0$), one arrives at the field theory~\cite{mccoy1978twodimensional,sachdev2011quantum}
\be
H_\text{ft} = \int_0^R \rd x \Big( \bar \psi \p_x \bar \psi - \psi \p_x \psi + im\bar \psi \psi + g \s \Big). \label{FT}
\ee
Here $R$ is the system size, $\bar \psi\,(\psi)$ are right (left) moving Majorana fermion fields, $m$ is the fermion mass ($m \sim 1 - h_x$), $g$ is the continuum longitudinal field, and $\s(x)$ is the spin operator in the continuum. For generic values of the parameters, both the lattice~\fr{lattice1D} and the continuum~\fr{FT} models are nonintegrable~\cite{onsager1944crystal,zamolodchikov1989integrals}. Herein we (mostly) focus on the ordered phase, $|h_x| < 1$ and $m > 0$. 
\begin{figure}
\includegraphics[width=0.38\textwidth,trim={0 0 0 5},clip]{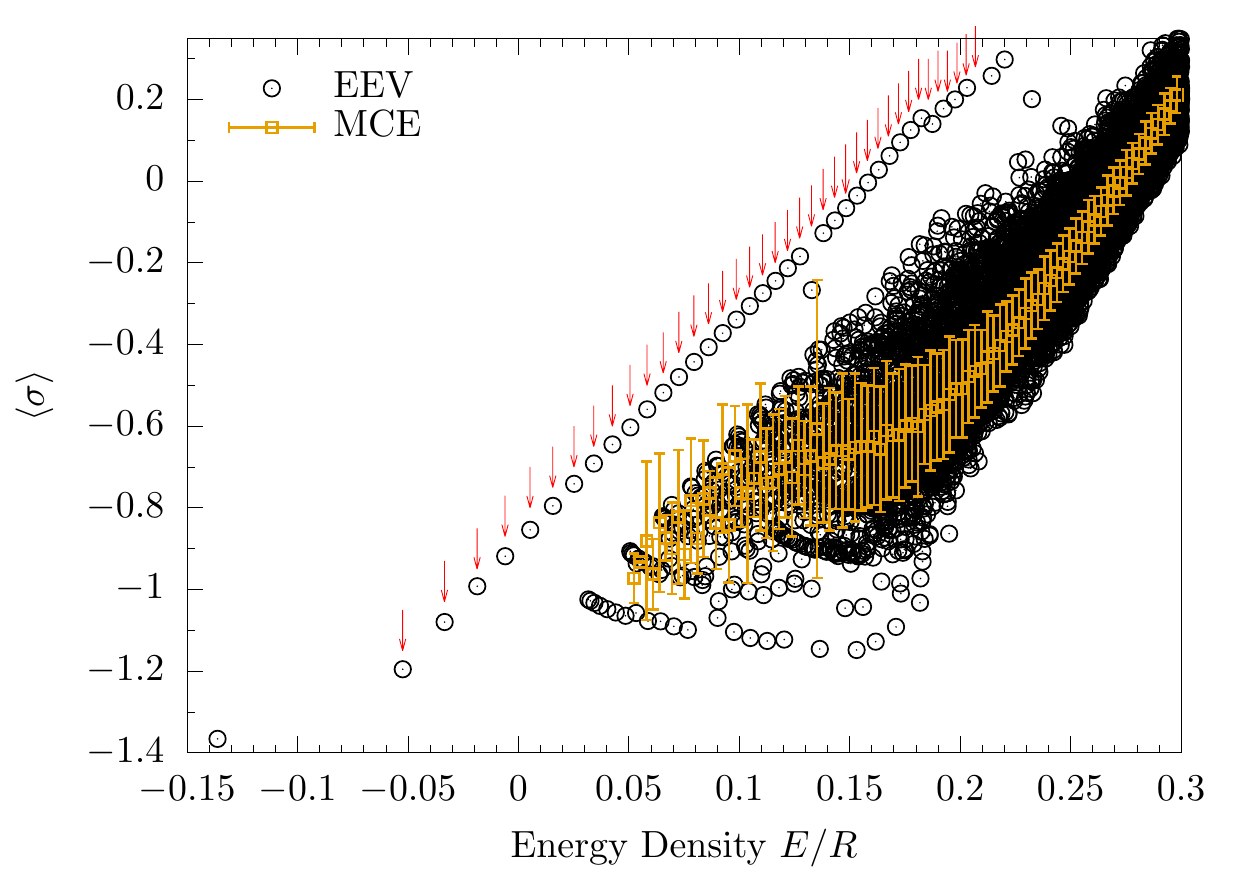} \\
\includegraphics[width=0.38\textwidth,trim={0 0 0 10},clip]{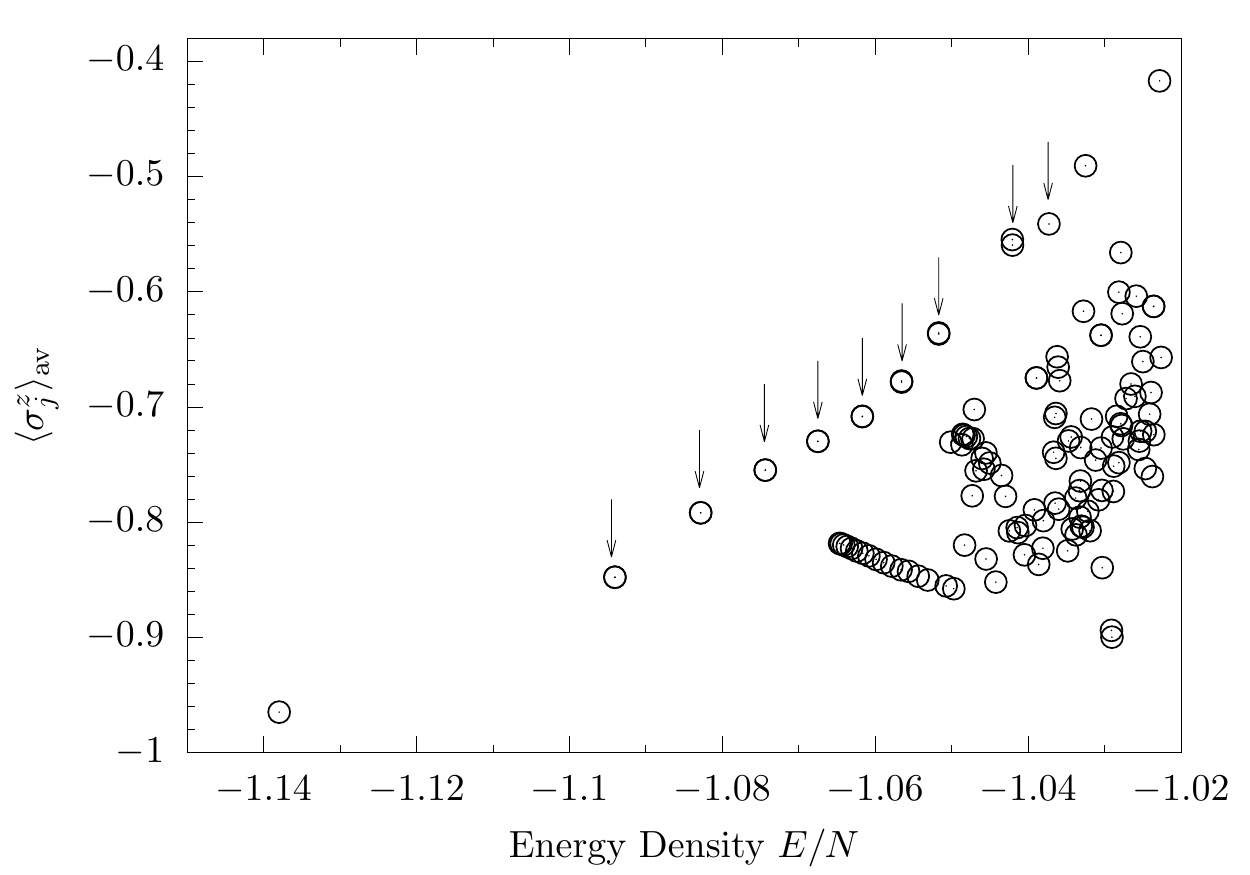}
\vspace{-2mm}
\caption{{\bfseries (Upper)} EEV spectrum of the spin operator $\s(0)$ as a function of energy $E$ for~\fr{FT} with $m=1$, $g=0.1$, $R=35$. Arrows show the (semiclassical) energies of the first forty meson states~\cite{fonseca2006ising}. The MCE (thermal) result is shown within the continuum, with error bars denoting the standard deviation of results averaged over. {\bfseries (Lower)} EEV spectrum of $\sum_j \s_j^z/N$ as a function of energy $E$ in the lattice model~\fr{lattice1D} with $J=-1$, $h_x = -0.5$, $h_z = 0.1$, for $N=40$ sites, computed with DMRG for open boundary conditions. Arrows are a guide for the eye, highlighting the nonthermal states.  Here the nonthermal states are meson like states confined to the vicinity of a boundary.}
\label{ising1d}
\end{figure}

In the absence of a longitudinal field ($h_z = 0$, $g=0$) low-energy excitations are spin flips (costing energy $\sim 2m$), which fractionalize into pairs of domain walls that are free to independently propagate. Thus, at low-energies, above energy $2m$ there is a continuum of two-particle states. The presence of a longitudinal field $h_z \neq 0$, $g \neq 0$ profoundly changes this. The energy cost of a domain of flipped spins now grows linearly in the size of the domain. This confining potential between domain walls (much like quarks in quantum chromodynamics (QCD)~\cite{sulejmanpasic2017confinement}) leads to a collapse of the low-energy continuum into discrete `meson' excitations, formed from pairs of domain walls~\cite{delfino1996nonintegrable,delfino1998nonintegrable}. This has been observed in two quasi-1D quantum magnets, CoNb$_2$O$_6$~\cite{coldea2010quantum,morris2014hierarchy} and SrCo$_{2}$V$_{2}$O$_{8}$~\cite{wang2015spinon,wang2016from}. 

The presence of confinement leads to nonthermal states appearing within the spectrum, despite the system being nonintegrable. To show this, we construct eigenstates of the two models,~\fr{lattice1D} and~\fr{FT}, and measure the average magnetization within each state~\footnote{That is, we compute the EV of $\s(0)$ in the field theory~\fr{FT} and $\sum_j \s^z_j/N$ in the lattice model~\fr{lattice1D}.}. On the lattice, we do this via the density matrix renormalization group (DMRG)~\cite{schollwock2011densitymatrix} by targeting up to 150 low-lying eigenstates~\footnote{For details of the DMRG procedure, see section S1.}. In the continuum we use truncated spectrum methods~\cite{james2018nonperturbative} to construct thousands of low-lying eigenstates~\footnote{For Fig.~\ref{ising1d} we use the truncated spectrum approach, with energy cut-off $E_\Lambda=10.5$, corresponding to constructing the Hamiltonian with the lowest $23,500$ basis states. The MCE was constructed by averaging over states within an energy window of size $\Delta E = 0.1$. We provide some further details of the truncated spectrum approach in Appendix B, and we also refer the reader to the recent review article~\cite{james2018nonperturbative}. We note that the truncation effects affect data on the right hand side of our plot, Fig.~\ref{ising1d}, above energy densities of $E/R \sim 0.2$. Further information about truncation effects, convergence, etc., is presented in Ref.~\cite{robinson2018signatures}.}. Example results are shown in Fig.~\ref{ising1d}; we see that there are two major features in the eigenstate expectation value (EEV) spectrum. Firstly, there is a thermal-like continuum of excitations on the right hand side of the plot (confirmed by comparison with the MCE in the continuum). With increasing system size, this continuum narrows as predicted by Eq.~\fr{eth}, see~\cite{robinson2018signatures}. Secondly, there is a line of states that is well separated and above this continuum (see the arrows in both plots), whose EVs do not coincide with the MCE results. These states remain separated from the thermal continuum up to the largest system sizes ($R\sim75$) that we can reach; extrapolation to the infinite volume limit is consistent with the nonthermal states possessing a different magnetization to the MCE, as shown in the supplemental. These features are seen in both the continuum and on the lattice; the similarity between the two panels in Fig.~\ref{ising1d} is striking.  

One advantage of tackling this problem in the continuum is that we have well-controlled analytical approaches, as well as the numerical data, that allow us to understand these nonthermal states. For example, in the upper panel we draw arrows at the energies of the meson (linearly-confined domain walls) excitations, as predicted from a semiclassical analysis~\cite{fonseca2003ising,rutkevich2005largen,fonseca2006ising}. We see that these coincide exactly with the nonthermal states. We also have direct access to the wave functions, and see that these states are well described by the two (domain wall) fermion sector of the theory~\cite{james2018nonperturbative}. The nonthermal states are well approximated by the meson form:
\be
| M_n\ra = \sum_{\nu=\text{NS},\text{R}} \sum_{p_\nu} \Psi_{n}(p_\nu) a\dg_{p_\nu}a\dg_{-p_\nu} |\nu\ra, 
\label{mesonState}
\ee
where $a\dg_{p_\nu}$ creates a fermion of momentum $p_\nu$ in the $\nu = \text{NS},\text{R}$ (Neveu-Schwarz and Ramond respectively) sector of the Hilbert space~\footnote{See Ref.~\cite{james2018nonperturbative} and the supplemental for further details.}, $|\nu\ra$ is the vacuum within a given sector. The wave function, $\Psi_n(p)$, and the mass of the meson, $M_n$, can be determined analytically via the Bethe-Salpeter equation, see~\footnote{See the supplemental for a brief overview of the derivation and solution of the Bethe-Salpeter equation for the meson eigenvalue problem. Reference~\cite{robinson2018signatures} also provides further details.} for details.

\noindent {\bf Meson stability above thresholds:}
The persistence of well-separated single meson excitations above the two-meson threshold is, at first glance, surprising. Analogously to QCD (see, e.g.,~\cite{sulejmanpasic2017confinement}), one might expect these single mesons to be unstable, with open decay channels to multimeson states.
As shown in Fig.~\ref{ising1d}, this intuition is incorrect. 
To shed some light on this, we consider the (two domain wall) meson excitations, described by~\fr{mesonState}, and compute the second order energy correction that comes from hybridization with four domain wall states and vacuum ~\footnote{We note that the zero temperature lifetime, $\tau$, i.e. the imaginary part of the self-energy,  has been computed by Rutkevich in Ref.~\cite{rutkevich2005largen} in the large $n$ (high meson energy) limit.}.  Below we will see that this correction is exceedingly small compared to the bound state energy, $E=M_n$, in contrast to the second order correction coming from the spin flip excitations within the disordered (paramagnetic) phase, where confinement is absent~\footnote{Note that we take the magnitude of the mass in the two phases to be identical, i.e. $-m_\text{disorder} = m _\text{order}$.}. We find that the meson corrections are orders of magnitude smaller than those of the paramagnetic spin flip. 

\begin{figure}
\includegraphics[width=0.35\textwidth]{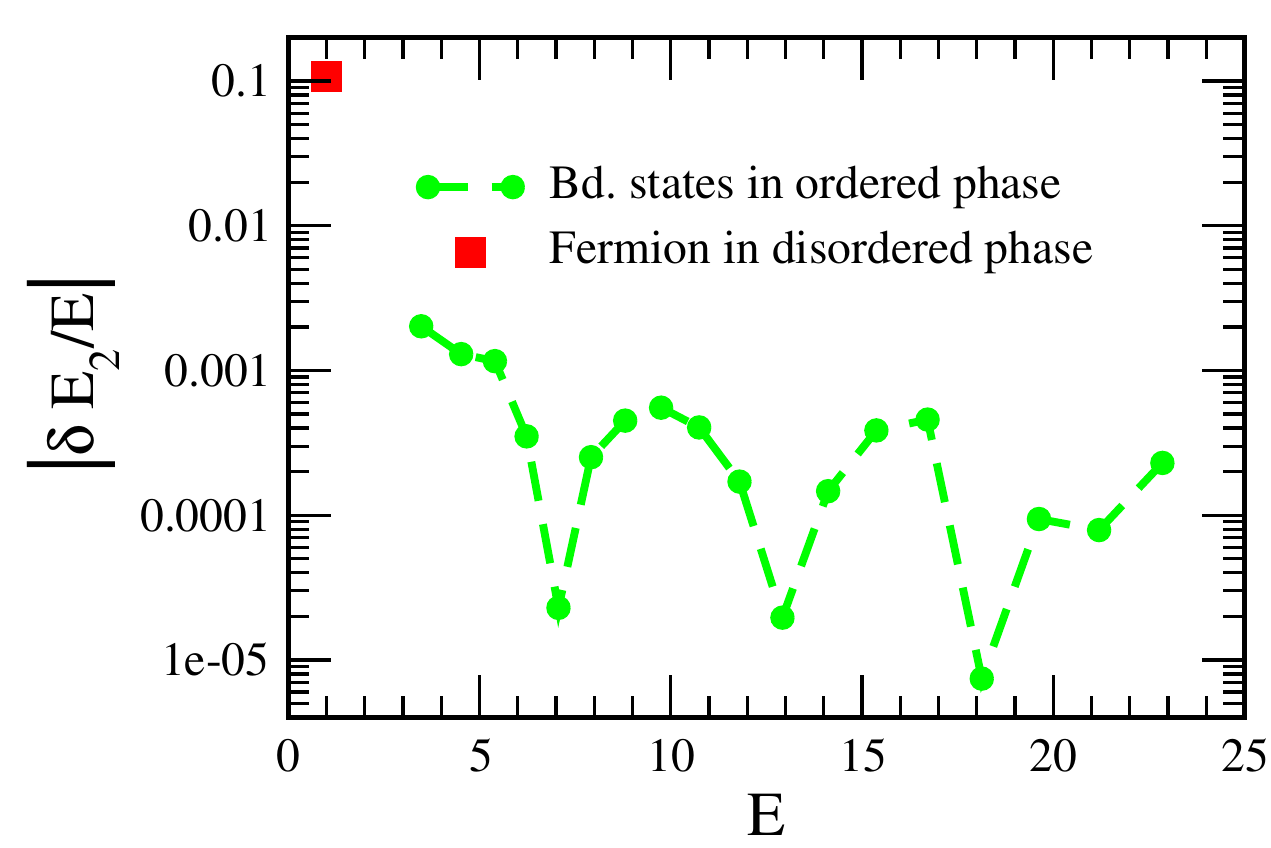} 
\vspace{-2mm}
\caption{We plot the relative second order correction to the energy of the first 19 zero momentum mesons ($\delta E_2/E$) coming from mixing with 0 and 4 domain wall states.  For comparison we also plot the energy correction to the zero momentum spin flip excitation in the disordered phase.  Note that the correction for the mesons is from $10^{-2}$ to $10^{-4}$ that of spin flip excitation.  Here $g=0.1$.}
\label{secondorderen}
\end{figure}

We give explicit details of the second order energy computation in the supplemental (see also~\cite{robinson2018signatures}), only schematically sketching the calculation here. 
The problem is split into three parts, $H = H_\text{meson} + H_\text{free} + H_\text{int}$. The first part, $H_\text{meson}$, describes the single meson part of the problem, whose eigenstates are given by~\fr{mesonState}. In $H_\text{free}$ we describe the non-interacting part of all the other ($n>2$) fermion sectors of the theory. Finally, $H_\text{int}$ contains all interaction vertices, except the two-fermion-to-two-fermion case, which was taken into account in $H_\text{meson}$.  We specifically consider vertices involve two-to-four fermions and two-to-zero fermions.  
A similar calculation is performed for the second order correction of the single particle excitations in the disordered phase. 

We present results of our computations in Fig.~\ref{secondorderen}, showing the relative second
order corrections to the zero momentum energy for the first nineteen meson states (green circles).  For comparison we present the corresponding
computation for the zero momentum fermion in the disordered phase (red square).  We see that the energy corrections
for all of the mesons range from $10^{-5}$ to $10^{-3}$ of their unperturbed energy.  Moreover the energy corrections 
for those mesons which lie above the four domain wall continuum, i.e. that are not nominally kinematically stable, are no
larger than those below the threshold.  We also see that the meson energy corrections are at least two orders of magnitude
smaller than the correction of the fermion (spin flip) for the disorder phase.
Thus the meson excitations, states of the form~\fr{mesonState}, appear to quasi-stable to mixing with four domain wall states.  This supports the results of the previous section; by slightly dressing the states~\fr{mesonState}, we form \textit{completely stable} nonthermal states in the finite volume.  Note that this is counter to the usual intuition from QCD, where one would expect the single meson to be unstable to kinematic decay above the two meson threshold.  
Even at higher orders in perturbation theory, the two domain wall sector of the theory appears to continue to mix only very weakly with the sectors containing $n\geq4$ domain walls,
despite there being scattering processes induced by the longitudinal magnetic field that remain finite into the thermodynamic limit.
While we have not extended our second order correction to account for mixing with six domain wall states explicity, we expect such mixing to be considerably smaller because of phase space considerations \cite{EsslerKonikReview}.

\noindent {\bf Extension to 2D:}
Surprisingly, the above analysis in 1D extends in a straightforward manner to higher dimensions. Consider the following 2D Hamiltonian:
\be
H_\text{2D} = \sum_{j} \Bigg( H_j + J_\perp \int_0^R \rd x \, \s_j(x) \s_{j+1}(x) \Bigg), 
\label{2D}
\ee
formed from individual Ising continuum chains
\be
H_j = \int_0^R \rd x \Big(  \bar\psi_j \p_x \bar \psi_j - \psi_j \p_x \psi_j + i m \bar\psi_j \psi_j  \Big),
\ee
coupled by a local spin-spin interaction of strength $J_\perp$.
For this system the coupling $J_\perp$ between neighbouring ordered ($m>0$) chains provides a confining potential.
Meson-like approximate eigenstates of~\fr{2D}, of the form
\be
|E_n\ra =  \sum_{\nu_k = \text{NS,R}}  \sum_{j=1}^N \sum_{p_{\nu_j}} \Psi_{n}^{\{\nu\}}(p_{\nu_j}) A\dg_j(p_{\nu_j})A\dg_j(-p_{\nu_j}) |\{\nu\}\ra, \label{mesonState2d}
\ee
can be found via an analogous Bethe-Salpeter equation~\footnote{As shown in the supplemental.}.
Here $N$ is the number of chains, $A\dg_j(p_\nu)$ creates a fermion in the $j$th chain with momentum $p_\nu$ in the $\nu=\text{NS,R}$ sector, and $|\{\nu\}\ra = \otimes_{j=1}^N |\nu_j\ra$ are the vacuum states of the system, formed from the individual $\nu_j$-vacua in each chain. The physical character of the wave function $\Psi_{n}^{\{\nu\}}(p_{\nu_j})$ is similar to the 1D case, Eq.~\fr{mesonState}. 

With meson states~\fr{mesonState2d} (i.e., approximate two fermion eigenstates) defined, one can proceed in a similar manner to the previous section, and compute their self-energies. This calculation is essentially identical to the previous case, leading us to conclude that meson excitations in 2D are extremely long-lived excitations. We can no longer construct the EEV spectrum in 2D (cf. Fig.~\ref{ising1d} in 1D), but a mean field decoupling of the 2D system into 1D chains suggests that these meson-like excitations should behave similarly to those analogous excitations in 1D, i.e. they are nonthermal states.
In the next section we provide further evidence of this.

\noindent {\bf Nonequilibrium dynamics in 2D:}
Having argued that nonthermal states exist in the 2D theory with confinement, Eq.~\fr{2D}, we now support this with evidence that the nonequilibrium dynamics is anomalous~\footnote{We note that we can easily show that there is an absence of thermalization in the 1D problem~\fr{FT}. A brief account of this is presented in the supplemental, with a detailed study presented in Ref.~\cite{robinson2018signatures}.}. This is one of the signatures of the presence of nonthermal states in the spectrum. Nonequilibrium dynamics is induced by a quench of the interaction $J_\perp=0  \to J_\perp \neq 0$. Both the initial state and subsequent time-evolution are computed in the chain array matrix product state (ChainAMPS) framework~\cite{james2018nonperturbative}. This methodology blends truncated spectrum methods with MPS algorithms, and has been used to study the entanglement entropy and spectrum of the 2D Ising model~\cite{james2013understanding}, and to compute the time-evolution following a quantum quench~\cite{james2015quantum}. 

\begin{figure}
\centering
\begin{tabular}{cc}
Ordered phase & Disordered phase\\
\includegraphics[width=0.24\textwidth]{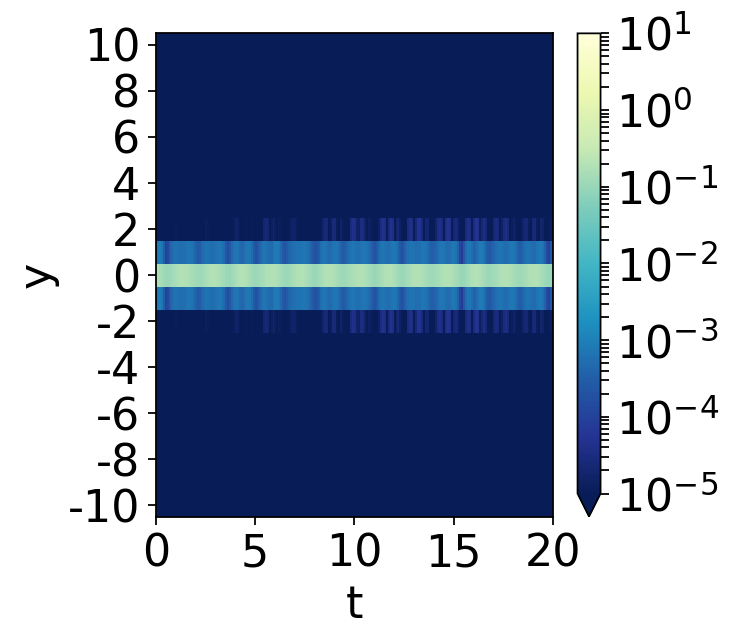}&\includegraphics[width=0.24\textwidth]{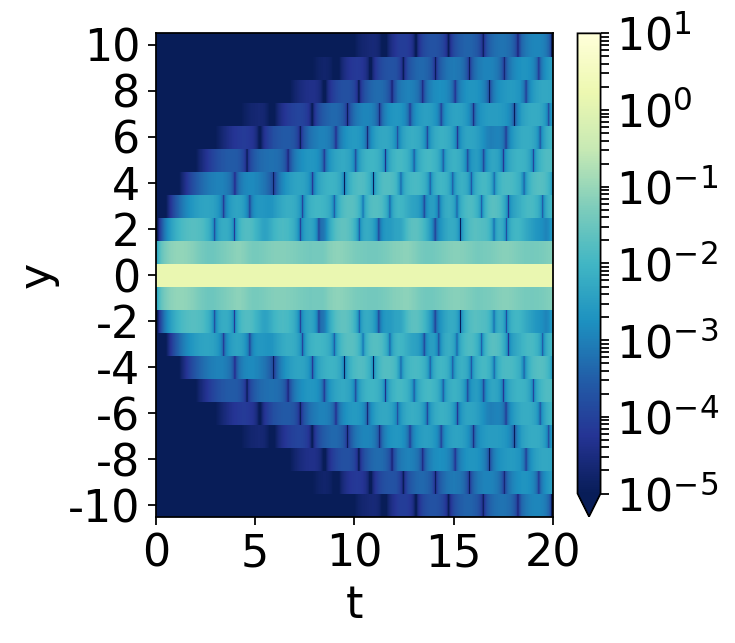} \\
\includegraphics[width=0.2\textwidth]{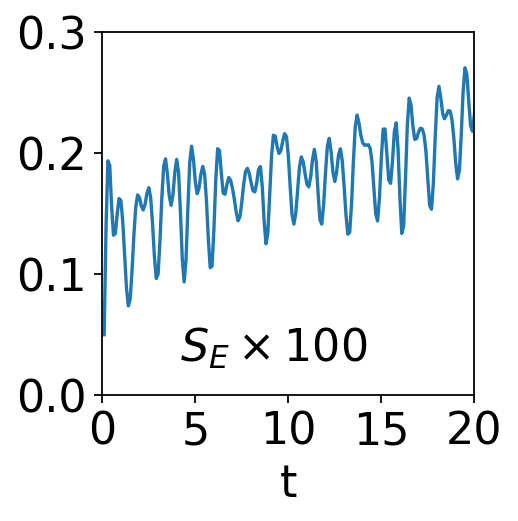} \hspace{0.15in} &\includegraphics[width=0.2\textwidth]{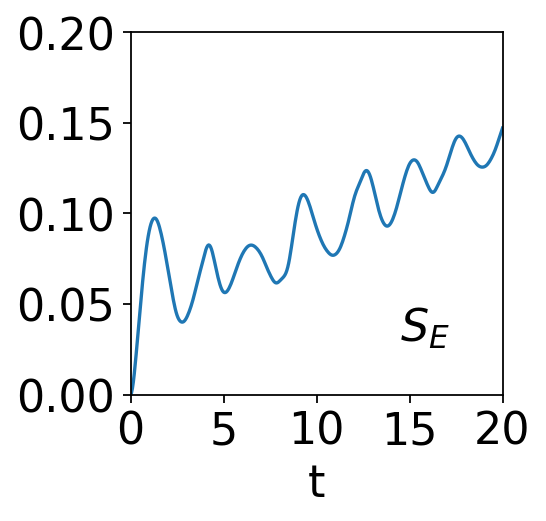}  \hspace{0.15in} 
\end{tabular}
\vspace{-4mm}
\caption{{\bfseries (Upper)} The time-evolution of the connected correlation function, $\left\vert\langle \sigma_{i+y}(x,t) \sigma_i(x,t) \rangle-\langle \sigma_{i+y}(x,t)\rangle\langle \sigma_i(x,t) \rangle\right\vert$, following quenches in the 2D quantum Ising model~\fr{2D} for $R=8, J_\perp=0$ to $-0.15$ and (left) ordered chains $m=1$, (right) disordered chains $m=-1$. Both cases start from a $J_\perp=0$ ground state. Dynamics are computed via ChainAMPS~\cite{james2018nonperturbative}. {\bfseries (Lower)} The time-evolution of the entanglement entropy $S_E$ following the same quenches. Note the $y$-axis of the lower left panel has been increased by a factor of 100. A detailed discussion of the simulations is provided in the Supplemental.}
\label{noneq2D}
\end{figure}

In Fig.~\ref{noneq2D} we present results for the time-dependence of the connected two-point spin correlation function between chains, $\left\vert\langle \sigma_{i+y}(x,t) \sigma_i(x,t) \rangle-\langle \sigma_{i+y}(x,t)\rangle\langle \sigma_i(x,t) \rangle\right\vert$, (upper panels) and the entanglement entropy $S_E$ (lower panels) for quenches from the $J_\perp=0$ ground state to $J_\perp\ne0$, for both ordered ($m>0$) and disordered ($m<0$) chains. Here $S_E$ is defined as the Von Neumann entanglement when the system is partitioned into two semi infinite arrays of chains. For ordered chains the two-point function does not show the usual light cone behavior following the quench, with response instead being strongly suppressed and correlations remaining local. In the presence of confinement this is consistent with the quasiparticle picture of Calabrese and Cardy~\cite{calabrese2006time,calabrese2007quantum}: the quench generates pairs of quasiparticles with opposing momenta (forming mesons in the presence of confinement), which propagate away from one another. At fixed energy density (as set by the quench), the particles can only separate a finite distance before the confinement potential saturates the available energy, and hence the light cone is suppressed.
In contrast, the disordered case, where confinement is absent, displays a clear light cone spread of correlations.
This suppression of the propagation of quasiparticles also impacts the growth of $S_E$ (with entanglement being carried by these quasiparticles), as is shown in the lower panels of Fig.~\ref{noneq2D}. 

Before concluding, we note that similar effects have been observed in the non-equilibrium dynamics of~\fr{lattice1D} and~\fr{FT}. In the lattice problem~\fr{lattice1D}, Kormos \textit{et al.}~\cite{kormos2017realtime} observed both a suppression of the light cone and the growth of the entanglement entropy following a global quantum quench. Non-equilibrium dynamics following quenches in the field theory~\fr{FT} have also shown clear signatures of the meson excitations~\cite{rakovsky2016hamiltonian,hodsagi2018quench,robinson2018signatures}.

\noindent {\bf Conclusions:}
In this Letter, we have seen that nonthermal states appear in the Ising model, in 1D and 2D, when confinement is present. The nonthermal states have EVs that do not match the MCE prediction, highlighting their nonthermal nature, despite an absence of integrability. We saw this very explicitly in Fig.~\ref{ising1d}, in both the continuum and on the lattice, by computing the EEV spectrum of the longitudinal magnetization.

We identified the nonthermal states as being meson-like, in that the state is well approximated by linearly confined pairs of domain walls, as expressed in Eqs.~\fr{mesonState} and~\fr{mesonState2d}. The mesons hybridize only very weakly with the thermal continuum of multimeson states, see Fig.~\ref{secondorderen}. From controlled numerical and analytical calculation in 1D, we turned our attention to 2D and argued that such meson states exist there, with essentially the same calculations applying in 1D and 2D. The presence of such nonthermal states can lead to anomalous nonequilibrium dynamics, illustrated in Fig.~\ref{noneq2D}, such as suppression of the lightcone and entanglement growth, as well as an absence of thermalization~\footnote{See the supplemental.} (for a recent example of this in a quantum quench of a 1D lattice model, see Ref.~\cite{2018arXiv180609674M}).

While we have focused on Ising models in 1D and 2D, it is natural to expect that the physics of nonthermal states carries over to other theories with confinement. Recently, holographic theories with confinement have shown an absence of thermalization~\cite{myers2017holographic}, a hallmark of the presence of nonthermal states. A natural test of this conjecture could be provided by the Schwinger model in an electric field, which has been the subject of a number of recent works~\cite{buyens2014matrix,buyens2016confinement,banuls2016chiral,banuls2017density,buyens2017finiterepresentation,buyens2017realtime} (the disordered Schwinger model has also recently been shown to display confinement driven non-ergodic behaviour \cite{nandkishore2017many,akhtar2018symmetry}).
We also note that recently kinetic constraints have been identified as a mechanism leading to non-ergodic eigenstates midspectrum \cite{turner2017quantum}.  These examples taken together suggest models possessing athermal eigenstates are not uncommon.  It will be a topic of future research to arrive at a classification scheme for such non-ergodic quantum systems. 

\acknowledgments{
\noindent {\bf Acknowledgments:}
All authors contributed equally to this work. We are grateful to Mari Carmen Ba\~nuls, Bruno Bertini, Pasquale Calabrese, Axel Cort\'es Cubero, Fabian Essler, Andrew Hallam, Dante Kennes, Andreas L\"auchli, Marcos Rigol, and Gabor T\'akacs for useful conversations. This work was supported by the Engineering and Physical Sciences Research Council, grant number EP/L010623/1 (A.J.A.J.), the U.S. Department of Energy, Office of Basic Energy Sciences, under Contract No. DE-SC0012704 (R.M.K.), and the European Union's Horizon 2020 research and innovation programme under grant agreement No 745944 (N.J.R.). We also acknowledge funding from the Simons Collaboration Programme entitled ``The Nonperturbative Bootstrap'' as part of the ``Hamiltonian methods in strongly coupled Quantum Field Theory'' meeting at the IHES Universit\'e Paris-Saclay, where part of this work was performed and presented.
}

\appendix

\setcounter{equation}{0}
\setcounter{figure}{0}
\setcounter{table}{0}
\makeatletter
\renewcommand{\theequation}{S\arabic{equation}}
\renewcommand{\thefigure}{S\arabic{figure}}

\section{S1: Details of the DMRG and additional data}
The spectrum of the lattice model, Eq.~\fr{lattice1D}, was calculated for an open chain of $N=40$ sites, by first using standard finite size DMRG techniques~\cite{schollwock2011densitymatrix} to find the ground state, and then a projector method to construct excited states (see Ref.~\cite{stoudenmire2012studying} for a nice description of this approach). This method is not guaranteed to find all the excited states in order, but if enough states are constructed and the weight parameter for the projector is chosen judiciously one can be reasonably confident of calculating the low energy spectrum correctly.
The maximum truncation error allowed in the DMRG algorithm was $10^{-10}$ and 20 finite size sweeps (20 left sweeps plus 20 right sweeps) were performed for each state.
To ensure that our DMRG routine was working appropriately, we carried out tests on a system of $N=14$ sites versus exact diagonalization.

In order to compare results from DMRG on a finite size system with open boundaries to results with periodic boundary conditions in the field theory~\fr{FT} (for which we consider only the zero momentum sector), we average the expectation values of the spin operator obtained in DMRG over all sites. This produces a zero (quasi)momentum quantity from the DMRG lattice simulations. We remind the reader that in the field theory~\fr{FT}, expectation values in zero momentum states $|k=0\ra$ satisfy 
\be
\la k = 0 | \sigma(x) | k=0 \ra = \frac{1}{R} \int_0^R \rd x\, \la k = 0 | \s(x) | k=0\ra, \nonumber
\ee
by translational invariance of the eigenstates. 

\begin{figure}
\includegraphics[width=0.8\linewidth]{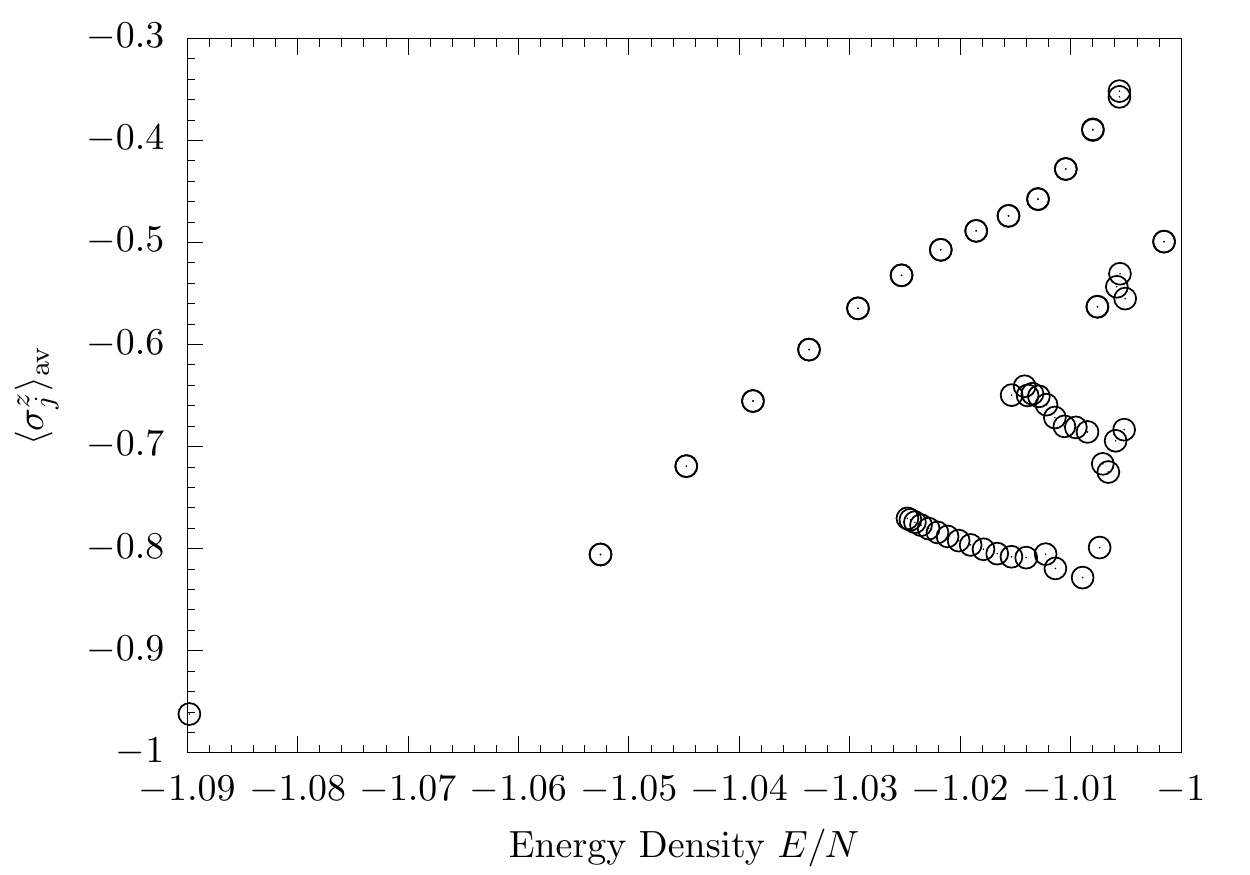} \\
\includegraphics[width=0.8\linewidth]{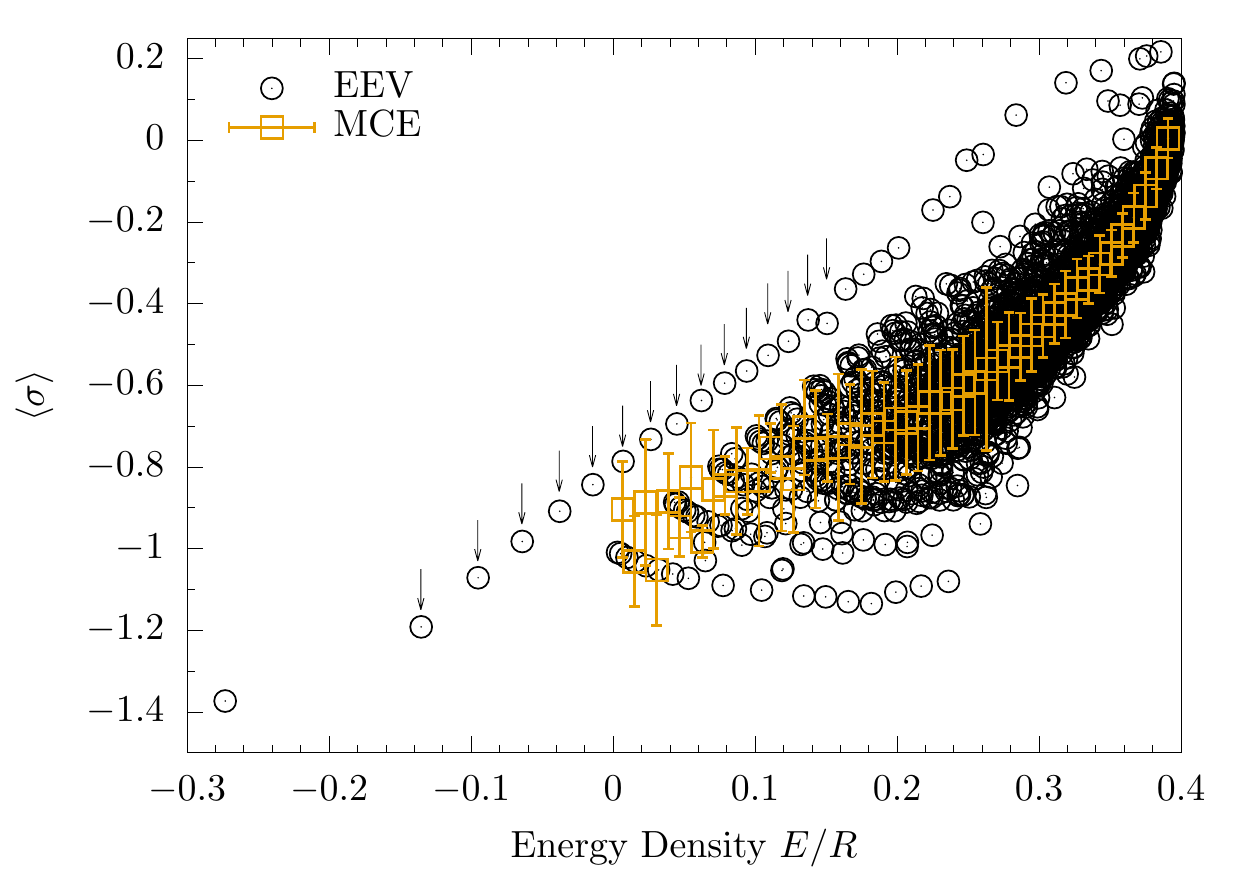}
\caption{{\bfseries (Upper)} Low-lying eigenstate expectation value spectrum of $\sum_j \s_j^z/N$ as a function of energy $E$ in the lattice model~\fr{lattice1D} with $J=-1$, $h_x = -0.5$, $h_z = 0.05$, for $N=40$ sites, computed with DMRG. {\bfseries (Lower)} A similar plot from the field theory~\fr{FT} with $m=1$, $g=0.2$ and $R=25$, for energy cutoff $E_\Lambda=12.5$. We highlight the first fifteen meson state energies, computed semiclassically~\cite{fonseca2006ising}, via arrows. See Fig.~\ref{ising1d} of the main text for similar data for $h_z = 0.1$, $g = 0.1$.}
\label{Fig:alt_lattice}
\end{figure}

In the upper panel of Fig.~\ref{Fig:alt_lattice} we present an alternative lattice data set for the smaller value of $h_z = 0.05$ (with all other parameters as in the lower panel of Fig.~\ref{ising1d}), showing that the nonthermal states are present in a range of parameters in the lattice model. Here the thermal continuum is slightly better converged than the results presented in the main body of the paper, showing clear structure that will broaden into the continuum in the infinite volume. Notice that again the nonthermal states persist for as far into the continuum as we can reach with DMRG, and have very well separated eigenstate expectation values. This is also true for the field theory~\fr{FT}, with the lower panel of Fig.~\ref{Fig:alt_lattice} showing comparable data for~\fr{FT} with $m=1$, $g=0.2$. Once again, the similarity between the lattice and continuum data is evident. 

\section{S2: Some details of the truncated space approach to the perturbed Ising field theory}

In this part of the supplemental, we recap the central details of the truncated spectrum approach to the perturbed Ising field theory. This was first developed by~~\textcite{yurov1991truncatedfermionicspace} and used to solve the eigenvalue problem for low-lying eigenstates.

We begin by considering a system of finite size $R$; for quantities such as the ground state energy this is expected to reproduce the infinite volume results up to corrections that are exponentially small in the system size (due to the presence of a spectral gap). We write the Hamiltonian in the form 
\be
H(m,g) = H_0(m) + gV,
\label{H0V}
\ee
with 
\begin{align}
H_0(m) &= \int_0^R \rd x \Big[ \bar \psi \p_x \bar \psi - \psi \p_x \psi + im\bar \psi \psi \Big], \\
V &= \int_0^R \rd x\ \s(x). 
\end{align}
Here $H_0(m)$ is the Hamiltonian for free fermions of mass $m$. The Hilbert space of $H_0(m)$ in the finite volume $R$ splits into two sectors, called Neveu-Schwartz (NS) and Ramond (R), which correspond to antiperiodic and periodic boundary conditions for the fermions~\cite{mccoy1978twodimensional}. 

Due to the difference in boundary conditions on the fermions in the two sectors of the Hilbert space, the momenta of the fermions within these sectors are quantized differently. The momenta of the fermions is quantized as $2\pi n/R$ with $n\in\mathbb{Z}+\frac12$ in the NS sector and $n\in\mathbb{Z}$ in the R sector. Eigenstates of $H_0(m)$ are obtained by acting on the vacuum with fermion creation operators:
\be
\begin{split}
& |k_1,\ldots,k_N\ra_{NS} = a\dg_{k_1}\ldots a\dg_{k_N} |0\ra_{NS},\quad k_i \in \mathbb{Z}+\frac12, \\
& |q_1,\ldots,q_N\ra_{R} = a\dg_{q_1}\ldots a\dg_{q_N} |0\ra_{R},\quad q_i \in \mathbb{Z},  \nonumber
\end{split}
\ee
Herein we use $k_i$ for the half-integer defining the momentum in the NS sector, and $q_i$ for the integer in the R sector. Creation operators obey the usual canonical anticommutation relations
\be
\{ a_{k}, a\dg_{k'} \} = \delta_{k,k'}, \quad \{ a_q , a\dg_{q'} \} = \delta_{q,q'}. \nonumber
\ee
An $N$-particle state is an eigenstate of $H_0(m)$ with energy ${\cal E}_N(R)$ given by
\bw
\be
{\cal E}_N(R)_\text{NS} = E_0(R)_\text{NS} + \sum_{i=1}^N \omega_{k_i}(R), \qquad 
{\cal E}_N(R)_\text{R} = E_0(R)_\text{R} + \sum_{i=1}^N \omega_{q_i}(R), 
\ee
where the single particle dispersion is $\omega_k(R) = \sqrt{m^2 + (2\pi k/R)^2 }$
and the vacuum energy is~\cite{mccoy1978twodimensional} 
\be
\begin{split}
E_0(R)_\text{NS} =& \frac{m^2 R}{8\pi}\log m^2 - |m| \int_{-\infty}^{\infty} \frac{\rd \theta}{2\pi} \cosh\theta \log \Big(1 + e^{-|m| R \cosh\theta}\Big), \\
E_0(R)_\text{R} =& \frac{m^2 R}{8\pi}\log m^2  - |m| \int_{-\infty}^{\infty} \frac{\rd \theta}{2\pi} \cosh\theta \log \Big(1 - e^{-|m| R \cosh\theta}\Big). \nonumber
\end{split}
\ee

The full problem~\fr{H0V} also features a finite strength of the magnetic field, $g$. The spin operator $\s(x)$ must also satisfy periodic boundary conditions in the finite volume, i.e. $\s(x+R) = \s(x)$. This restricts the free fermion states allowed in each sector of the Hilbert space, and this restriction depends on the sign of the mass $m$ (i.e., the phase of the Ising model)~\cite{mccoy1978twodimensional}:
\be
m > 0~:~\begin{array}{l} 
\text{NS-states with}~N_f\in2\mathbb{Z},\\
\text{R-states with}~N_f\in2\mathbb{Z}.
\end{array}
\qquad
m < 0~:~\begin{array}{l} 
\text{NS-states with}~N_f\in2\mathbb{Z},\\
\text{R-states with}~N_f\in2\mathbb{Z}+1. 
\end{array}
\ee
Here $N_f$ is the number of fermions within the basis states. 

As well as modifying the allowed states, the spin operator in the Hamiltonian~\fr{H0V} couples the states in different sectors of the Hilbert space. The matrix elements of the spin operator, in the basis of eigenstates of $H_0(m)$, are known from integrability~\cite{bugrij2000correlation,bugrij2001form,fonseca2003ising}
\be
{}_{NS}\la k_1,\ldots,k_K| \s(0,0) | q_1,\ldots,q_N\ra_R = S(R) \prod_{j=1}^K \tilde g(\theta_{k_j}) \prod_{i=1}^N g(\theta_{q_i}) F_{K,N}(\theta_{k_1},\ldots,\theta_{k_N}|\theta_{q_1},\ldots,\theta_{q_N}), \nonumber
\ee
where $\theta_{k_n}$ are finite-size rapidities that satisfy $|m|R \sinh\theta_{k_n} = 2\pi k_n$ (and similar for $q_n$), 
\be
g(\theta) = \frac{e^{\kappa(\theta)}}{\sqrt{|m|R\cosh\theta}}, \qquad \tilde g(\theta) = \frac{e^{-\kappa(\theta)}}{\sqrt{|m|R\cosh\theta}}, \quad \kappa(\theta) = \int_{-\infty}^\infty \frac{\rd \theta'}{2\pi} \frac{1}{\cosh(\theta-\theta')} \log\left( \frac{1 - e^{-|m|R\cosh\theta'}}{1 + e^{-|m|R\cosh\theta'}}\right), \nonumber
\ee
$F_{K,N}$ is the matrix element of the spin-field in the infinite volume~\cite{berg1979construction}
\be
F_{K,N}(\theta_1,\ldots,\theta_K|\theta_1',\ldots,\theta_N') =  
i^{\left\lfloor \frac{K+N}{R} \right\rfloor}\bar\s \prod_{0<i < j \leq K} \tanh\left( \frac{\theta_i - \theta_j}{2}\right) 
\prod_{0<p<q\leq N} \tanh\left( \frac{\theta'_i - \theta'_j}{2}\right) 
\prod_{0 < s \leq K} \prod_{0 < t \leq N} \coth \left(\frac{\theta_s - \theta'_t}{2} \right),
\label{FEqn}
\ee
where $\bar\s = \bar s |m|^{1/8}$, $\bar s = 2^{1/12} e^{-1/8} A^{3/2} = 1.35783834\ldots$, and $A$ is Glashier's constant. The factor $S(R)$ is the vacuum expectation value:
\be
\bar \s S(R) = \left\{ 
\begin{array}{lcl} 
{}_{NS} \la 0| \s(0,0) |0\ra_R & \quad & \text{for } m > 0, \\
{}_{NS} \la 0| \mu(0,0) |0\ra_R & \quad & \text{for } m < 0, 
\end{array} \right.   \nonumber
\ee 
where $\mu(0,0)$ is the disorder parameter (dual to $\s(0,0)$) in the Ising field theory~\fr{FT}. This expectation value also has a known form, due to Sachdev~\cite{sachdev1996universal}
\be
S(R) = \exp\Bigg(\frac{(mR)^2}{2} \iint_{-\infty}^{\infty} \frac{\rd \theta_1 \rd \theta_2}{(2\pi)^2} \frac{\sinh\theta_1 \sinh\theta_2}{\sinh(mR\cosh\theta_1) \sinh(mR\cosh\theta_2)} \log \left| \coth \frac{\theta_1-\theta_2}{2} \right| \Bigg). \nonumber
\ee
\ew

With the above knowledge at hand, we can form the complete Hamiltonian matrix~\fr{H0V} in the basis of free fermion eigenstates. Of course, we cannot deal with the full Hamiltonian, as the Hilbert space is still infinite. So, to tackle the problem numerically the Hilbert space is truncated. This truncation is motivated by the renormalization group relevant properties of the perturbing operator $\s(x)$. With scaling dimension $1/8$ the operator is strongly relevant, and as a result it strongly mixes low energy basis states. However, \textit{it does not strongly couple low and high energy basis states}, and so one can truncate the basis through the introduction of a cutoff.

There are numerous ways to truncated the basis. For example, in Ref.~\cite{fonseca2003ising} the authors propose a simple scheme, where the `level' of a state is defined through
\be
|p_1,\ldots,p_N\ra : \quad \text{level } \equiv \ell = \frac12 \sum_i |p_i|.  \nonumber
\ee
and the Hilbert space is truncated at some maximal level $L = \ell_\text{max}$. Symmetries of the model, such as translational invariance, can be implemented by working in a given symmetry sector (e.g., at fixed momentum $P = \sum_i p_i$). Following such a truncation, the Hamiltonian is a finite matrix and can be diagonalized with standard numerical routines. In the Ising theory, such a procedure has been shown in many cases to produce results in excellent agreement with other analytical or numerical approaches, see e.g. Ref.~\cite{fonseca2003ising}. 

Extensions to mitigate the Hilbert space truncation are possible, see the recent review~\cite{james2018nonperturbative} for further details. A detailed discussion of convergence, etc., is given in Ref.~\cite{robinson2018signatures}.

\section{S3: Finite-size scaling of the nonthermal states}

\begin{figure}
\includegraphics[width=0.8\linewidth]{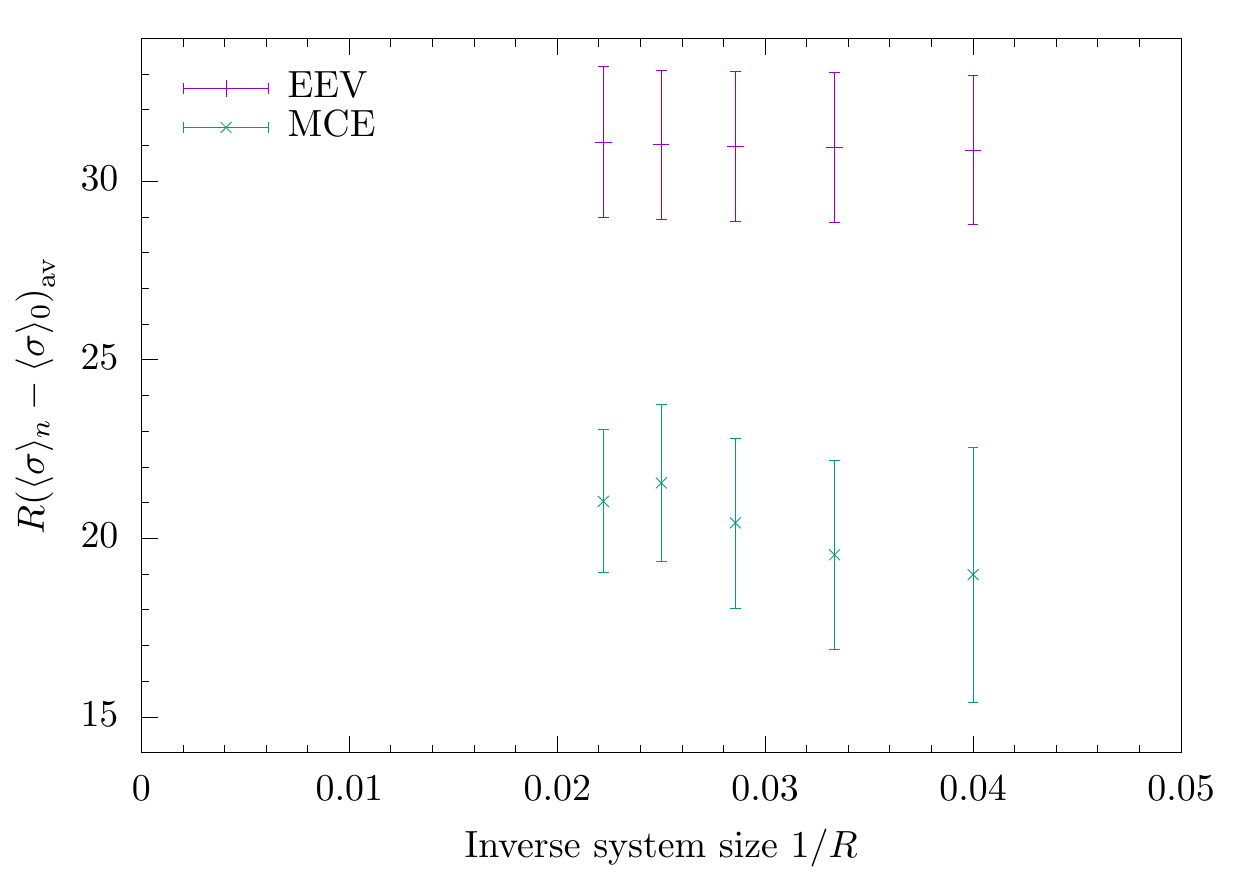}
\caption{The average magnetization $R\la \s(0)\ra$ of the $n=11\to15$ nonthermal states, measured relative to the ground state $R\la \s \ra_0$, for $g=0.1$. We compare this to the corresponding averaged microcanonical ensemble (MCE) prediction, showing clear separation between the magnetization of the meson states and that of the thermal continuum at the same energy, which persists to the infinite volume limit. Error bars show the standard deviation of data averaged over.}
\label{finitesizescaling}
\end{figure}

In this section, we consider the nonthermal states that appear in the eigenstate expectation value (EEV) spectrum of the field theory~\fr{FT}, and how they vary as a function of system size $R$ in comparison to the microcanonical ensemble (MCE, thermal) prediction. For brevity, we focus on the case presented within the main text, the field theory~\fr{FT} with $m=1$ and $g=0.1$, and consider the EV of the integrated spin operator $R\la \s(0)\ra$ (i.e. the magnetization of the state). An example of the EEV spectrum for the \textit{local} magnetization with system size $R=35$ is presented in the upper panel of Fig.~\ref{ising1d} of the main text, where the nonthermal states are highlighted by arrows. 

We present our example results in Fig.~\ref{finitesizescaling}. For clarity we consider the average magnetization of the $n=11\to15$ nonthermal states, measured relative to the ground state. These states have energies well within the multiparticle continuum, cf. Fig.~\ref{ising1d}, and we can construct them to high precision up to systems of size $R\sim50$. We compare this average magnetization to the corresponding thermal prediction from the MCE. It is apparent that the magnetization of the nonthermal states \textit{is not consistent with the thermal prediction, even in the infinite volume limit, $1/R\to0$}. 

\section{S4: Bethe-Salpeter equation for the meson wave function}

Let us now consider the meson excitations, both in the 1D theory~\fr{FT} and 2D problem~\fr{2D}. They are described, respectively, by the wave functions (see Eqs.~\fr{mesonState} and~\fr{mesonState2d})
\be
\begin{split}
|M_n\ra &= \sum_{\nu=\text{NS,R}} \sum_{p_\nu} \Psi_n(p_\nu) a\dg_{p_\nu} a\dg_{-p_\nu} |\nu\ra, \\
|E_n\ra &= \sum_{\nu_k=\text{NS,R}}\sum_{j=1}^N \sum_{p_{\nu_j}} \Psi^{\{\nu\}}_{n,j}(p_{\nu_j}) A\dg_j(p_{\nu_j})A\dg_j(-p_{\nu_j}) |\{\nu\}\ra. 
\end{split}
\label{mesons}
\ee 
In this part of the supplemental we determine, through relevant Bethe-Salpeter equations, the functional form of the wave functions, $\Psi_n(p)$ and $\Psi_n^{\{\nu\}}(p_\nu)$. In the 1D case we will be relatively terse, presenting the detailed calculation in Ref.~\cite{robinson2018signatures}. 

\subsubsection{S4.i. The 1D perturbed Ising theory~\fr{FT}}

Our analysis here is along the same lines as Fonseca and Zamolodchikov~\cite{fonseca2003ising}; the Bethe-Salpeter equation for the meson wave function $\Psi_n(p_\nu)$ is derived by restricting the Schr\"odinger equation to the manifold of meson-like states. To obtain this, we have to evaluate the matrix elements $\la \pm p_v | H |M_n \ra$ where $|\pm p_v\ra \equiv a\dg_{p_v} a\dg_{-p_v}|v\ra$ are two fermion states. The resulting Schr\"odinger equation reads [cf. Eq.~\fr{mesons}] 
\be
\begin{split}
M_n\Psi_{n}(p_v) &= 2 \omega(p_v)\Psi_{n}(p_v) \\
&+ \frac{gR}{2} \sum_{v',q_{v'}} \Psi_{n}(q_{v'}) \la \pm p_v | \s(0) | \pm q_{v'}\ra.
\end{split}
\label{resSE}
\ee
where we define the free fermion dispersion relation
$\omega(p) = \sqrt{p^2+m^2}$. The matrix elements of the spin operator $\s(0)$ on free fermion states are known (see Refs.~\onlinecite{bugrij2000correlation,bugrij2001form,fonseca2003ising}) and only connects states in different sectors $v = \text{NS},\text{R}$ of the Hilbert space, hence $v' = \bar v$ ($v=\text{NS},\text{R}$, $\bar v = \text{R},\text{NS}$). In the large but finite volume $R$ we have 
\be
\begin{split}
&\sum_{q_{\bar v}}  \la \pm p_v | \s(0) |\pm q_{\bar v} \ra \Psi_{n}(q_{\bar v}) = \\
&- \sum_{q_{\bar v}} \frac{\bar \s}{R^2} \frac{p_v q_{\bar v}}{\omega(p_v)^2\omega(q_{\bar v})^2} \left( \frac{\omega(p_v) + \omega(q_{\bar v})}{\omega(p_v)-\omega(q_{\bar v})}\right)^2 \Psi_{n}(q_{\bar v}).
 \end{split}
\label{ME}
\ee

Following some manipulations, detailed in Ref.~\cite{robinson2018signatures}, Eq.~\fr{resSE} can be cast into the form 
\be
\epsilon_n \Psi_n(y) = \Big(|y| -  \p_y^2 - \frac{t^2}{4} \p_y^4 - \frac{t^4}{8} \p_y^6 \Big) \Psi_n(y) - t^4 \delta'(y) \Psi'_n(0),
\label{BSEq} 
\ee
where $y = xmt$ with $t = (g\bar\s/m^2)^{1/3}$, $\Psi_n(y)$ is the (real space) wave function of the $n$th meson excitation, $\epsilon_n m t^2 = (M_n - 2m + g\bar\s R/2)$, and we have kept only the leading small momentum terms in an expansion of the free fermion dispersion. We call Eq.~\fr{BSEq} the Bethe-Salpeter equation.  

Taking $t$ to be small (i.e. for small longitudinal field $g$) , Eq.~\fr{BSEq} can be solved perturbatively. Solutions take the form 
\be
\Psi_n(y) = {\rm sgn}(y) F_n(|y|-\epsilon_n), 
\label{propsol}
\ee
where $F_n(y)$ is a solution of the homogenous equation 
\be
0 = \Big(y - \p_y^2 - \mu t^2 \p_y^4 - \nu t^4 \p_y^6\Big) F(y),
\label{BSgenred}
\ee
with $\mu = 1/4$, $\nu = 1/8$. These solutions can, in turn, be written in terms of solutions $A(y)$ of Airy's equation, 
\be
(y - \p_y^2)A(y) = 0, \label{airyeq}
\ee
reading 
\be
F_A(y) = A(y) + t^2 F_A^{(2)}(y) + t^4 F_A^{(4)}(y)+ \ldots,  \nonumber
\ee
We give explicit expressions for the prefactors $F_A^{(n)}$ for $n=2,4,6$ in Ref.~\cite{robinson2018signatures}. 

The solution~\fr{propsol} will satisfy Eq.~\fr{BSEq} provided the following boundary conditions are fulfilled 
\be
\begin{split}
&(i)~~~~F_n(-\epsilon_n) = O(t^2), \\
&(ii)~~~ \mu F_n(-\epsilon_n) + vt^2 F''_n(-\epsilon_n) = O(t^4), \\
&(iii)~~ F_n(-\epsilon_n) + \mu t^2 F''_n(-\epsilon_n) + v t^4 F^{(4)}_n(-\epsilon_n)\\
&\qquad  - \frac{1}{2} t^4 F'_n(-\epsilon_n) = O(t^6). 
\end{split}
\label{BCs}
\ee
The boundary conditions~\fr{BCs} allow us to fix the particular form of $F_n(y)$ 
\be
F_n(y) = F_{{\rm Ai}}(y) + \alpha_n(\epsilon_n) F_{{\rm Bi}}(y). \nonumber
\ee
Here ${\rm Ai}(y)$, ${\rm Bi}(y)$ are the two linearly independent solutions to Airy's equation~\fr{airyeq}. The function $\alpha_n(\epsilon_n)$ can also be written as a power series in $t$:
\be
\alpha_n(\epsilon_n) = \alpha_{0,n}(\epsilon_n) + t^2 \alpha_{2,n}(\epsilon_n) + t^4 \alpha_{4,n}(\epsilon_n) + O(t^6).  \nonumber
\ee
with $\alpha_{i,n}(\epsilon_n)$ being fixed by the boundary conditions~\fr{BCs}, see~\cite{robinson2018signatures}.

To complete the solution of Eq.~\fr{BSEq}, we restrict our attention to normalizable $\Psi_n(y)$. Using that $\lim_{y\to\infty} {\rm Bi}(y) = \infty$, this forces us to find $\epsilon_n$ such that $\alpha_n(\epsilon_n)=0$. Combining this condition with the above, we arrive at
\be
\epsilon_n = - z_n + \delta_{2,n} t^2 + \delta_{4,n} t^4 + O(t^6), \nonumber
\ee
where ${\rm Ai}(z_n) = 0$, and 
\be
\begin{split}
&\delta_{2,n} = - \frac{\mu}{5} z_n^2, \\
&\delta_{4,n} = \bigg( \frac{84\mu^2}{350} - \frac{2\mu^2}{25} - \frac{\nu}{7}\bigg) z_n^3 - \bigg( \frac{2\mu^2}{5} - \frac{4\nu}{7} + \frac{1}{2}\bigg).
\end{split}
\label{defdelta}
\ee
To recover our Bethe-Salpeter equation for the meson wave function, we set $\mu = 1/4$ and $\nu = 1/8$. The meson energy, $M_n$, can then be expressed as a simple power series in $t$ 
\be
M_n - E_0 =  2m \Big( 1 + a_{2,n} t^2 +  a_{4,n} t^4 + a_{6,n} t^6 + O(t^8) \Big).  \label{En2}
\ee
Here the ground state energy is $E_0 = -m^2Rt^3/2$, and we have the following dimensionless parameters
\be
a_{2,n} = - \frac{z_n}{2}, \quad a_{4,n} = -\frac{z_n^2}{40}, \quad
a_{6,n} = -\frac{127}{560} - \frac{11z_n^3}{2800}. \nonumber
\ee
The meson energies~\fr{En2} agree with previous calculations by other authors~\cite{fonseca2003ising,fonseca2006ising}.

\subsubsection{S4.ii. The 2D coupled chain problem~\fr{2D}}

An analogous calculation to the 1D case can be performed in 2D~\fr{2D} to obtain the meson wave function. The only significant difference in this case is that one has to carefully keep track of the different Hilbert space sectors on each chain. The meson states in 2D takes the form given in Eq.~\fr{mesons}, where
\be
|\{\nu\}\ra = |\nu_1\ra \otimes |\nu_2\ra \otimes \ldots \otimes |\nu_N\ra
\ee
is a vacuum state of the system, formed from the tensor product of vacuum states on each chain, $\nu_i \in \{ \text{R}, \text{NS}\}$,  and the wave function carries vacuum indices
\be
\Psi^{\{\nu\}}_{n,j}(p_{\nu_j}) \equiv \Psi_{n,j}^{\{\nu_1,\ldots,\nu_N\}}(p_{\nu_j}).
\ee
The restricted Schr\"odinger equation analogous to Eq.~\fr{resSE} reads
\bw
\be
\begin{split}
E \Psi^{\{\nu\}}_{n,j}(p_{\nu_j}) =& 2\omega(p_{\nu_j}) \Psi^{\{\nu\}}_{n,j}(p_{\nu_j}) + u R \bar\s^2 \sum_{l \neq j-1,j} \Psi^{\{\nu_1,\ldots, \bar \nu_l, \bar \nu_{l+1}, \ldots, \nu_N \}}_{n,j}(p_{\nu_j}) \\
& + \frac{uR}{2} \la \mp p_{\nu_j} | \s_j(0) | \bar \nu_j\ra \sum_{q_{\bar\nu_{j+1}}} \la \nu_{j+1} |\s_{j+1}(0)|\mp q_{\bar\nu_{j+1}}\ra  \Psi^{\{\nu_1,\ldots, \bar \nu_j, \bar \nu_{j+1}, \ldots, \nu_N \}}_{n,j+1}(q_{\bar\nu_{j+1}}) \\
& + \frac{uR}{2} \sum_{q_{\bar \nu_{j-1}}} \la \nu_{j-1}| \s_{j-1}(0) | \mp q_{\bar \nu_{j-1}} \ra \la \mp p_{\nu_j} | \s_j(0) | \bar \nu_j\ra  \Psi^{\{\nu_1,\ldots, \bar \nu_{j-1}, \bar \nu_{j}, \ldots, \nu_N \}}_{n,j-1}(q_{\bar\nu_{j-1}}) \\
& + \frac{\bar\s u R}{2} \sum_{q_{\bar\nu_j}} \la \mp p_{\nu_j} | \s_i(0) | \mp q_{\bar\nu_j} \ra \Big(  \Psi^{\{\nu_1,\ldots, \bar \nu_{j-1}, \bar \nu_{j}, \ldots, \nu_N \}}_{n,j}(q_{\bar\nu_j}) +  \Psi^{\{\nu_1,\ldots, \bar \nu_j, \bar \nu_{j+1}, \ldots, \nu_N \}}_{n,j}(q_{\bar \nu_j}) \Big).
\end{split}
\label{2dresse_track}
\ee
Here we define $\bar \nu_j = \text{R},\text{NS}$ when $\nu_j = \text{NS},\text{R}$ and $\bar \s = \la \nu_j | \s(0) | \bar\nu_j\ra$, as in previous sections. 

In Eq.~\fr{2dresse_track} we have kept careful track of the vacuum states on each of the chains. We now make an assumption about the wave function, which is justified a posteriori by the energy of the bound states containing a term that gives the correct ground state energy. We assume
\be
 \Psi^{\{\nu_1,\ldots, \bar \nu_l, \bar \nu_{l+1}, \ldots, \nu_N \}}_{n,j}(p_{\nu_j}) = \left\{
 \begin{array}{lll}
 - \Psi^{\{\nu_1,\ldots, \nu_l, \nu_{l+1}, \ldots, \nu_N \}}_{n,j}(p_{\nu_j}) & \quad & \text{if~} l \neq j-1,j , \\
  \Psi^{\{\nu_1,\ldots, \nu_l, \nu_{l+1}, \ldots, \nu_N \}}_{n,j}(p_{\nu_j}) & \quad & \text{if~} l = j-1,j. 
  \end{array}\right.
\ee 
This assumption allows us to simplify notations $ \Psi^{\{\nu\}}_{n,j}(p_{\nu_j}) \equiv \Psi_{n,j}(p_{\nu_j})$ and Eq.~\fr{2dresse} becomes 
\be
\begin{split}
E \Psi_{n,j}(p_{\nu_j}) =& 2\omega(p_{\nu_j}) \Psi_{n,j}(p_{\nu_j}) - u R \bar\s^2 (N-2) \Psi_{n,j}(p_{\nu_j}) \\
& + \frac{uR}{2} \la \mp p_{\nu_j} | \s_j(0) | \bar \nu_j\ra \sum_{q_{\bar\nu_{j+1}}} \la \nu_{j+1} |\s_{j+1}(0)|\mp q_{\bar\nu_{j+1}}\ra  \Psi_{n,j+1}(q_{\bar\nu_{j+1}}) \\
& + \frac{uR}{2} \sum_{q_{\bar \nu_{j-1}}} \la \nu_{j-1}| \s_{j-1}(0) | \mp q_{\bar \nu_{j-1}} \ra \la \mp p_{\nu_j} | \s_j(0) | \bar \nu_j\ra  \Psi_{n,j-1}(q_{\bar\nu_{j-1}}) \\
& + \bar\s u R \sum_{q_{\bar\nu_j}} \la \mp p_{\nu_j} | \s_i(0) | \mp q_{\bar\nu_j} \ra  \Psi_{n,j}(q_{\bar \nu_j}).
\end{split}
\label{2dresse}
\ee

Following a sequence of steps similar to the 1D problem, Eq.~\fr{2dresse} can be recast in the form 
\be
\begin{split}
\epsilon_n  \Psi_{n,j}(y) &= \Big( |y|  - \p_y^2 - \frac{s^2}{4} \p^4_y - \frac{s^4}{8} \p^6_y \Big) \Psi_{n,j}(y) + \frac{s^4}{8} \delta'(y) \Big( 4 \Psi'_{n,j}(0) + \Psi'_{n,j+1}(0) + \Psi'_{n,j-1}(0) \Big), 
\end{split}
\label{BS2d}
\ee
\ew
where $s = (4\bar \s^2 u/m^2)^{1/3}$, $y = msx$ and $ms^2 \epsilon_n = E_n - 2 m + NuR \bar\s^2$. Here, as with the 1D problem, we have kept only the leading terms in a power series (in momentum) expansion of the free fermion dispersion. Eq.~\fr{BS2d} is the Bethe-Salpeter equation for the 2D problem. 

Fourier transforming the wave function in the interchain direction
\be
\tilde\Psi_{n,q}(y) = \sum_{j=1}^N e^{ijq} \Psi_{n,j}(y),  
\ee
we arrive at an equation of the form 
\be
\begin{split}
\epsilon_n \tilde\Psi_{n,q}(y) =& \Big( |y| - \p_y^2 - \mu s^2 \p^4_y - \nu s^4 \p_y^6 \Big) \tilde \Psi_{n,q}(y)\\
&+ \rho_q s^4 \delta'(y) \tilde \Psi_{n,q}'(0). 
\end{split} 
\ee
where
\be
\mu = \frac14, \quad \nu = \frac18, \quad \rho_q = \frac14 \cos(q) + \frac12. 
\ee
Here we see we have arrived at a slight modification of Eq.~\fr{BSEq}, obtained in the 1D problem. This equation can then be solved in a similar manner to the previous section. The solution has energy [cf. Eqs.~\fr{defdelta}]
\be
\begin{split}
\epsilon_n &= -z_i + \bar \delta_2 s^2 + \bar \delta_4 s^4 + \ldots, \\
\bar \delta_2 &= - \frac{\mu}{5} z_i^2, \nn
\bar\delta_4 &= \bigg( \frac{84\mu^2}{350} - \frac{2\mu^2}{25} - \frac{\nu}{7} \bigg)z_i^3  
- \bigg( \frac{2\mu^2}{5} - \frac{4\nu}{7} + \frac{\rho_q}{2} \bigg).
\end{split}
\ee

\section{S5: Computing the self-energy of meson excitations}
In the previous section of the supplemental, we derived and solved the Bethe-Salpeter equation for the meson wave functions in the 1D and 2D problems. These states, Eqs.~\fr{mesons}, are \textit{not exact eigenstates} of~\fr{FT} (although they well-approximate exact eigenstates) and so these two-fermion mesons are in principle susceptible to mixing with 2n-fermion states (where here $n=0,4,6\cdots$).
Here we compute the second order energy correction that comes from this mixing as a parameterization of its strength and hence
the likelihood the meson will survive as a distinct state above kinematic thresholds where it can nominally decay.
A detailed exposition will be given in Ref.~\cite{robinson2018signatures}. Here we only give the essential overview of the problem.  

We begin by separating the Hamiltonian into a part treated exactly (whose eigenstates are the meson states) $H_0$ and a part that we can treat perturbatively. This separation reads $H_\text{pert}$: 
\begin{eqnarray}
H &=& H_0 + H_\text{pert},  \nn
H_0 &=& \sum_{n=0}^\infty H_{2n,0} + g \int \rd x\, \mathbb{P}_2 \s(x) \mathbb{P}_2, \label{mesonH0}\\
H_{2n,0} &=& \sum_{\nu \in \text{NS}, \text{R}}  \sum_{k_\nu} \omega(k_\nu) \mathbb{P}_{2n} a\dg_{k_\nu} a_{k_\nu} \mathbb{P}_{2n}, \label{mesonH2n} \\
H_\text{pert} &=& \sum_{\substack{n,m\\ (n,m) \neq (2,2)}} g \int \rd x \, \mathbb{P}_n \s(x) \mathbb{P}_m, \label{mesonHpert}
\end{eqnarray}
where $\mathbb{P}_n$ is the projector onto the $n$-particle part of the Hilbert space, and $a\dg_{p_\nu}$ creates a fermion of momentum $p_\nu$ in
sector $\nu = \text{NS}, \text{R}$.

Written in this form, $H_0$~\fr{mesonH0} contains the full theory restricted to the two-particle subspace (i.e., $\mathbb{P}_2 H \mathbb{P}_2$), as well as the noninteracting part of all other sectors, Eq.~\fr{mesonH2n}. In the previous section of the supplemental, we solved the two-particle problem, obtaining the meson states. The part that we will treat perturbatively, $H_\text{pert}$, contains all terms that couple mesons to states with other numbers of particles, as well as the interactions between states with higher particle numbers. 

We can further rewrite the two particle part of the problem by defining operators $b\dg_{ak}$ that create the $a$th meson with momentum $k$. Then $H_0$ reads: 
\be
H_0 = \sum_a \sum_k E_a(k) b\dg_{ak}b_{ak} + H_{0,0} + \sum_{n=2}^\infty H_{2n,0}. \nonumber
\ee
Here $E_a(k)$ is the dispersion relation for the meson created by $b\dg_{ak}$.  In terms of the Fermi operators, we write $b\dg_{ak}$ as
\begin{eqnarray}
b\dg_{ak} &=& \sum_{\nu=\text{NS,R}}b\dg_{\nu ak}\cr\cr
b\dg_{\nu ak} &=& \sum_{p_\nu} \Psi_a(p_\nu) a^\dagger_{p_\nu}a^\dagger_{-p_\nu},
\end{eqnarray}
where we have divided $b\dg_{ak}$ explicitly into its Ramond and Neveu-Schwarz parts.

\begin{widetext}
The second order energy correction to the mesons that comes from mixing with 0 and 4 domain wall states is then given by (to ${\cal O}(g^2)$)
\begin{eqnarray}
\delta E_{2i} &=& \frac{1}{M_i}\sum_{\nu = \text{R},\text{NS}}|f^{2,\nu}_{a}|^2 + \sum_{\nu = \text{NS,R}}\sum_{q_{\nu1}<q_{\nu 2}<q_{\nu 3}<q_{\nu 4}} \delta_{0,\sum q_{\nu i}} \frac{|f^{4,\nu}_{a,q_{\nu 1},q_{\nu 2}, q_{\nu 3},q_{\nu 4}}|^2}{M_i - \omega(q_{\nu 1}) - \omega(q_{\nu 2}) - \omega(q_{\nu 3}) - \omega(q_{\nu 4})},
\end{eqnarray}
where
\begin{eqnarray}
f^{2,\nu}_{a} &=& g\bar\sigma
\sum_{q_{\bar\nu}} \Psi_a(q_{\bar\nu}) \frac{S(R)}{R\omega(q_{\bar\nu})}F_{2,0}(\theta_{q_{\bar\nu}},-\theta_{q_{\bar\nu}});\cr\cr
f^{4,\nu}_{a,q_{\nu 1},q_{\nu 2},q_{\nu 3},q_{\nu 4}} &=& gR\bar\sigma\sum_{q_{\bar \nu}} \Psi_a(q_{\bar\nu})\frac{S(R) F_{2,4}(\theta_{q_{\bar\nu}},\theta_{-q_{\bar\nu}}|\theta_{q_{\nu 1}}, \theta_{q_{\nu 2}}, \theta_{q_{\nu 3}}, \theta_{q_{\nu 4}})}{R^3 \big(\omega(q_{\bar\nu})
\omega(-q_{\bar\nu})\prod_{i=1}^4\omega(q_{\nu i})\big)^{1/2}},
\end{eqnarray}
and $F_{K,N}$ is defined in Eq.~\fr{FEqn}.

\end{widetext}

This correction is expected to take the form
\begin{equation}
\delta E_{2i} = \alpha R + \beta_i,
\end{equation}
i.e. it will have a term scaling with the volume $R$. This is to be expected as the unperturbed ground state energy of the system $\delta E_{2gs}$ will pick up a volume term from being allowed to mix with the $2n$-domain wall states.  It is also the case that the correction proportional to $R$ for the ground state must be exactly the same as that of the meson states (the energy difference between a meson excitation and the ground state cannot scale linearly with $R$).  It is, however, difficult
to exhibit this explicitly because it requires summing over the second order correction coming from states involving arbitrary domain wall number.  (Happily in the corresponding computation for the spin flips in the paramagnetic phase, we can analytically exhibit this $R$-dependence and demonstrate that it does indeed cancel.)  Here we thus compute (numerically) $\delta E_{2i}$ as a function of $R$, fit the resulting dependence, and extract the constant $\beta_i$ term.  It is this that is plotted in Fig.~\ref{secondorderen}.

\section{S6: Computation of self-energy of spin flips in paramagnetic phase}

We can perform a similar computation of the second order energy correction of a spin flip excitation of the quantum Ising model in its disordered paramagnetic phase (i.e. the energy correction of a single Ramond fermion).  The point here is to compare the overall magnitudes of the energy corrections in the ordered and disordered cases. Here the leading contribution is a process by which the spin flip scatters into two spin flips or into a state with no spin flips:
\begin{widetext}
\begin{eqnarray}
\delta E_{2sf} &=& \frac{g^2R^2\bar\sigma^2}{2}\sum_{q\in \text{NS}}\frac{1}{mR^3\omega(q)^2\big(m-2\omega(q)\big)^2}\frac{\tanh^2(\theta_q)}{\tanh^4(\theta_q/2)} + \frac{g^2R\bar\sigma^2}{m^2}\cr\cr
&=& \alpha_{sf}R + \beta_{sf}
\end{eqnarray}
\end{widetext}
We are able here to extract $\alpha_{sf}$ exactly here as follows,
\begin{eqnarray}
\alpha_{sf} &=& -\frac{g^2\bar\sigma^2}{m^2},
\end{eqnarray}
while we obtain $\beta_{sf}$ numerically.
We can similarly compute the correction to the ground state energy in the paramagnetic phase coming from mixing with the
single flip sectors:
\begin{eqnarray}
\delta E_{2gs} = -\frac{g^2R\bar\sigma^2}{m^2}
\end{eqnarray}
We see that in calculating the difference $\delta E_{2sf}-E_{2gs}$, the volume terms proportional to $R$ exactly cancel.  It would be interesting to see if this cancellation can be exhibited explicitly for contributions involving a larger number of spin flips. In Fig.~\ref{secondorderen}, $\beta_{sf}$ is plotted as a red square.

\section{S7: Some details of the ChainAMPS simulations}
ChainAMPS~\cite{james2018nonperturbative} constructs a 2D quantum system by coupling together an array of chain models with truncated spectra. Each chain acts as a `super'-site on a 1D lattice, and this specialized 1D model can be treated using standard matrix product state techniques, including time evolution algorithms~\cite{james2015quantum}.
For our simulations we consider a system of infinitely many chains, initially in a $J_\perp=0$ (uncoupled) ground state, corresponding to a product state of chain ground states.
Specifically for the ordered chains case ($m=1$) we took a product state of the symmetric superposition of the Ramond and Neveu-Schwarz vacuum states for each chain (strictly these are degenerate only in the $R\to \infty$ limit, though the energy difference is exponentially small in $R$). The spectrum of each chain (of length $R=8$) was truncated to include the lowest 167 states. Using the notation NS(R)$X$ to label Neveu-Schwarz(Ramond) states of $X$ fermions these are : 2 vacua (NS0 and R0), 66 R2 states, 64 NS2 states, 21 R4 states and 14 NS4 states. 
This system was evolved under the Hamiltonian Eq.~\fr{2D} with $J_\perp=-0.15$ for $t>0$, using infinite time evolving block decimation (iTEBD)~\cite{vidal2007classical} with a 2nd order Trotter decomposition and time step size of $0.02$ and a bond dimension $\chi=100$ so that the truncation error at each step was of order $10^{-10}$ at the longest times.

For comparison, in Fig.~\ref{noneq2D}, we also show a quench in the disordered phase ($m=-1$ ), with chains of length $R=8$ and an initial state that is a product state of Neveu-Schwarz chain vacuum states.
In this case the lowest 163 states were kept: 1 NS0, 17 R1, 64 NS2, 67 R3 and 14 NS4.
The post quench interchain coupling was again $J_\perp=-0.15$ and a 2nd order Trotter decomposition with time step size of $0.01$ was used in the time evolution, with a larger bond dimension $\chi=200$ so that the truncation error at each step was still of order $10^{-10}$ at the longest times.

\section{S8: Nonthermalizing initial states in 1D quenches}

In this final section of the supplemental, we consider how nonthermal states affect the non-equilibrium dynamics of the 1D system, Eq.~\fr{FT}. We do so through an illustrative example; a detailed discussion and study can be found in Ref.~\cite{robinson2018signatures}. The hallmark of thermalization in a nonequilibrium context is the agreement of the diagonal ensemble (DE) with the microcanonical ensemble (MCE) constructed at the appropriate energy density~\cite{rigol2008thermalization,rigol2009breakdown}.  

\begin{figure}
\includegraphics[width=0.4\textwidth]{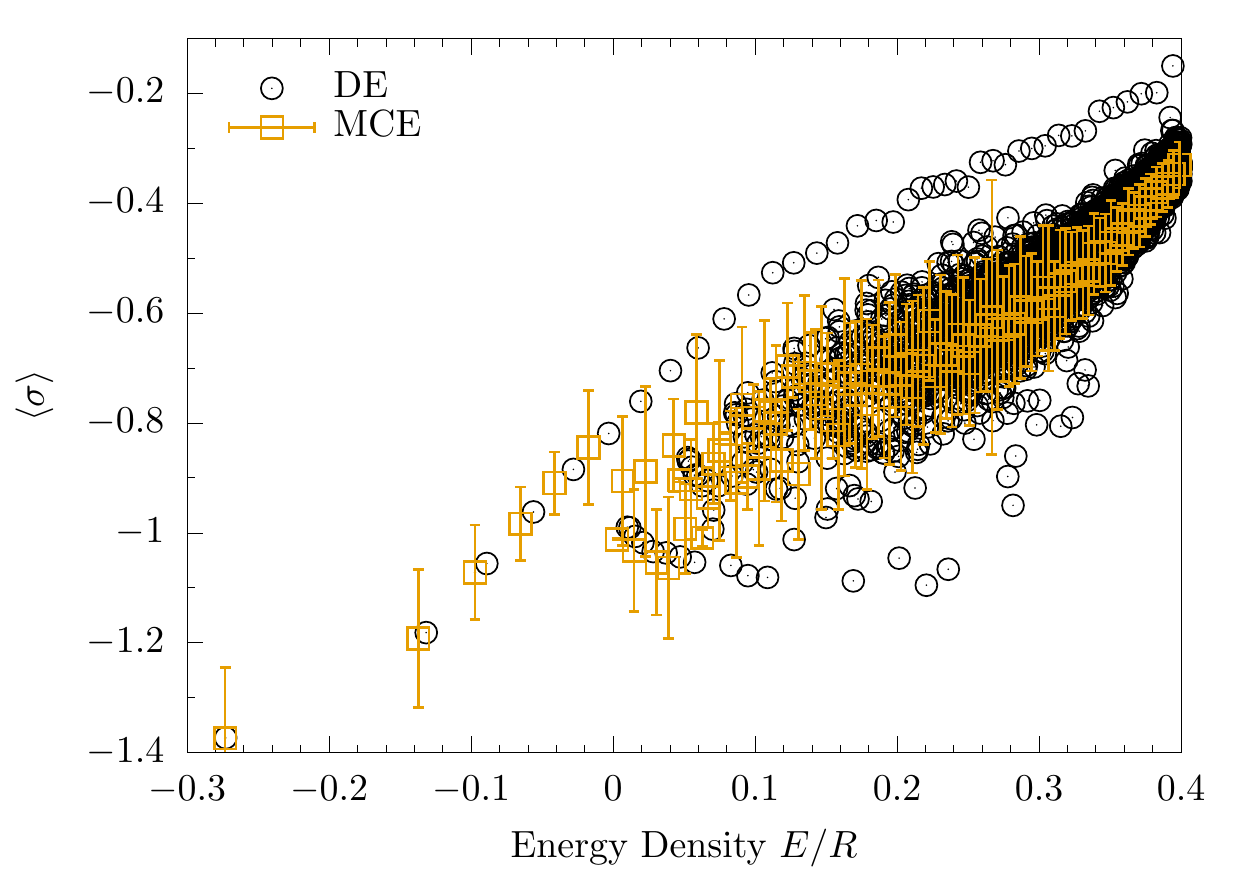}
\caption{A comparison of the DE and MCE predictions for the long-time limit of $\la\s\ra$ following a quantum quench. We start from eigenstates of~\fr{FT} with $m=1$ and $g=0.1$, subsequently time-evolving with $m=1$, $g=0.2$. Here the system size $R=25$ and we used an energy cutoff of $E_\Lambda = 13.5$.}
\label{demce}
\end{figure}

In Fig.~\ref{demce} we show a comparison between the DE and MCE predictions following a quantum quench, where we start in eigenstates of~\fr{FT} with $g=0.1$ (constructed with truncated spectrum methods) and time-evolve according to~\fr{FT} with $g=0.2$. We see that this comparison looks very similar to the EEV spectrum shown in Fig.~\ref{ising1d} of the main text! Many states project strongly onto the nonthermal states present within the spectrum, leading to a band of states with nonthermal EVs in the long time limit. We also see, however, that almost all starting states thermalize, with the MCE and DE predictions for EVs coinciding. 

\begin{figure}
\includegraphics[width=0.4\textwidth]{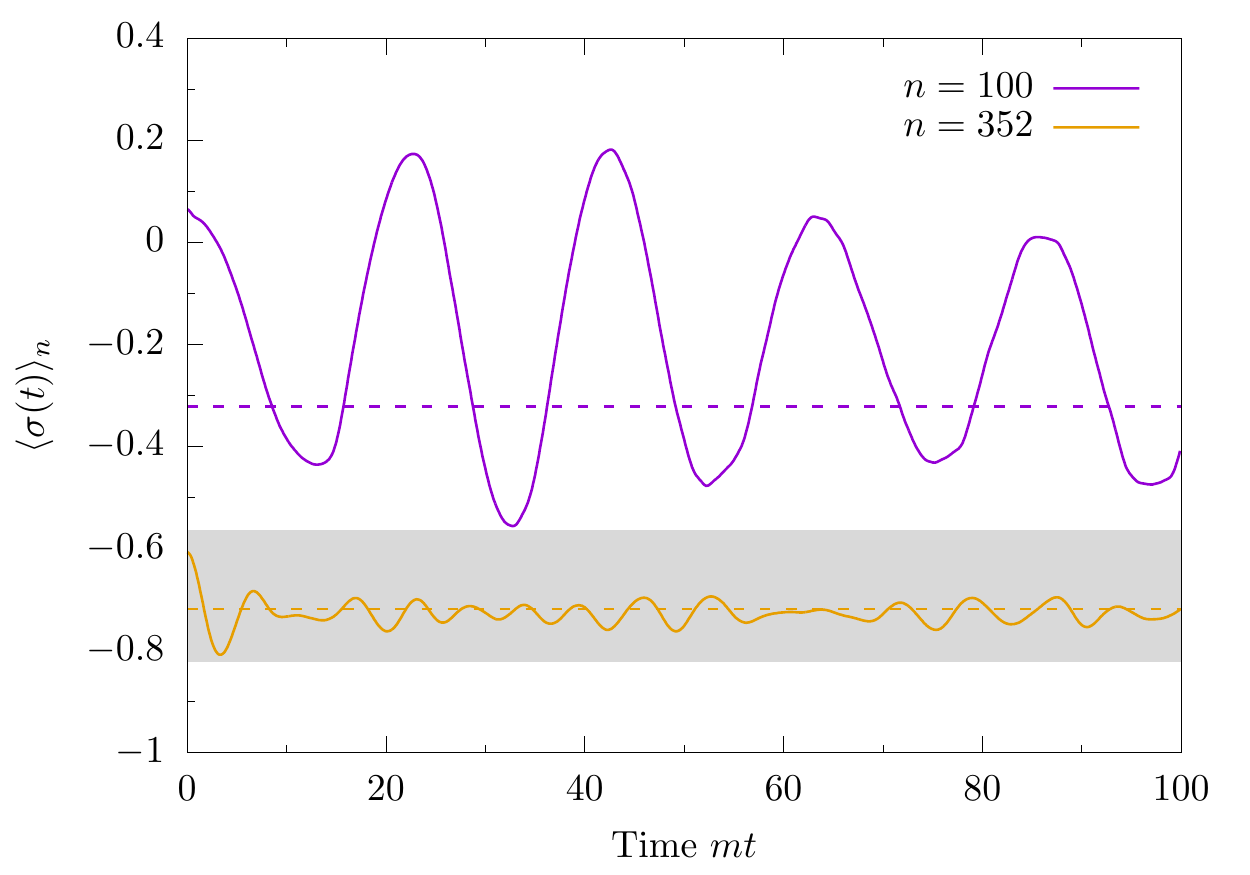}
\caption{Time-evolution of the local magnetization $\s(0)$ following a quantum quench $g=0.1 \to g=0.2$ for a nonthermalizing state ($n=100$) and a thermalizing state ($n=352$) with similar energy densities $E/R\sim0.208$. Predictions for the long time limit from the DE are shown via dashed lines, and the result of the MCE (plus the standard deviation) is denoted by the shaded region, cf. Fig.~\ref{demce}.}
\label{timeevo}
\end{figure}

Differences between nonthermalizing and thermalizing states can also be observed in the real time dynamics of observables (i.e., not just in the long time limit). In Fig.~\ref{timeevo} we compare the real-time dynamics of the local magnetization following the quench $g=0.1\to0.2$ when starting from two initial states of (almost) the same energy density, $E/R \sim 0.208$. The first initial state (the $n=100$ eigenstate of~\fr{FT} with $g=0.1$) is nonthermalizing: the DE result does not agree with the MCE prediction. On the other hand, the second state ($n=352$) is thermalizing with the long-time limit of the EV described well by the MCE. 

We see that there are significant differences between the real time dynamics of the thermalizing and nonthermalizing initial states. In the first case, the thermalizing state rapidly decays to and fluctuates about its long-time, thermal, value. In contrast, the nonthermalizing state exhibits slowly decaying large amplitude oscillations about its nonthermal DE result.

\bibliography{RareStatesBib}

\end{document}